\documentstyle[pra,aps,twocolumn,graphics,epsfig]{revtex}

\begin{document}
\draft \title{The Born and Markov approximations for atom lasers.} 
\author{G.M. Moy \cite{email:Moy}, J.J. Hope and C.M. Savage} 
\address{Department of Physics and Theoretical Physics, The Australian 
National University, \\
Australian Capital Territory 0200,
Australia.}
\date{\today}
\maketitle

\begin{abstract}
We discuss the use of the Born and Markov approximations in describing the 
dynamics of an atom laser.  In particular, we investigate the applicability 
of the quantum optical Born-Markov master equation for describing output 
coupling.  We derive conditions based on the atomic reservoir, and atom 
dispersion relations for when the Born-Markov approximations are valid and 
discuss parameter regimes where these approximations fail in our atom laser 
model.  Differences between the standard optical laser model and the atom 
laser are due to a combination of factors, including the parameter regimes 
in which a typical atom laser would operate, the different reservoir state 
which is appropriate for atoms, and the different dispersion relations 
between atoms and photons.  We present results based on an exact method in 
the regimes in which the Born-Markov approximation fails.  The exact 
solutions in some experimentally relavent parameter regimes give 
non-exponential loss of atoms from a cavity.

\end{abstract}


\narrowtext

\section{Introduction} \label{sec:Intro}

An atom laser is a device which, in analogy to the optical laser, 
would emit a coherent beam of bosonic atoms.  Various models of atom 
lasers have been presented 
\cite{Holland95,Wiseman95a,Wiseman95b,Guzman96,Spreeuw95,Olshanii95,Moy97,Wiseman97}. 
 
Many of these schemes are based around a master equation description 
for the dynamics of the 
system\cite{Holland95,Wiseman95a,Wiseman95b,Guzman96,Wiseman97}.  In 
these, the atom laser output is described by a Born-Markov master 
equation.  Making the Born-Markov approximation involves assuming that 
reservoir correlations decay rapidly compared with the system damping 
time and that the reservoir does not change significantly with time 
due to the effect of the system.  We discuss these approximations in 
the context of atom output coupling.

Born-Markov master equations were initially derived for optical 
systems\cite{QuantumOptics}.  There they are used to describe a system 
(for instance an optical laser mode) coupled to a large, unchanging 
reservoir (the external modes).  In optics the Born-Markov 
approximations allow an equation containing only system variables to 
be derived.  One of the fundamental assumptions involved in deriving 
such a Born-Markov master equation is that the reservoir correlation 
functions decay rapidly.  This allows the reservoir to be approximated 
as unchanging in time.  While it is assumed the system does not affect 
the reservoir, the reservoir does affect the system.  

An equation in terms of system variables alone is also an important 
goal for describing atom lasers.  However, atom and photon devices 
work in substantially different parameter regimes.  Moreover, atoms 
and photons have different dispersion relations, which affects the 
behaviour of the reservoir correlation functions.  Furthermore, 
typically the reservoir for an optical cavity is taken to be at 
thermal equilibrium at a nonzero temperature.  For atoms, a vacuum 
reservoir with all modes initially empty is often more appropriate.

In this paper we look at the validity of the Born-Markov master 
equation for describing output coupling from a single mode atomic 
system to a reservoir.  The reservoir is described as a continuum of 
free-space modes.  Our description of the coupling will initially be 
quite general, though our later discussion will focus on particular 
schemes in which the atom becomes free of the system through a change 
of state.  Such a change of state to an untrapped state can be 
achieved using either a Raman transition \cite{Moy97} or an RF 
transition\cite{MITExpt,MITExpt2,MIT3}.  We will also discuss the 
effects of gravity on output coupling.  In the latter part of the 
paper we will focus on broadband coupling.  This allows a comparison 
with exact results\cite{Hope97,Moy97b}, however we also present finite 
coupling results which more accurately describe output coupling 
through a change of state.  We will discuss exact equations which can 
be obtained in regimes where the Born-Markov approximations fail.

In section \ref{sec:Output} we present our model of output coupling 
from a single mode trap.  This allows exact equations of motion, and 
their solutions, to be obtained for system variables.  In section 
\ref{sec:MasterEqn} we model this system using the master equation 
formalism, emphasising the importance of the Born and Markov 
approximations.  In section \ref{sec:Validity} we then consider the 
validity of these approximations.  In the optical case, the adequacy 
of the Born-Markov approximation means that the standard model we are 
using leads to exponential decay of the number of photons in a cavity.  
The equivalent model for atoms does not lead to an atom number which 
tends to zero for sufficiently long time.  In section 
\ref{sec:nonzero} we discuss the reasons for this and consider the 
results obtained from models which include the effects of gravity.  
Having noted that the Born-Markov approximations may fail, we discuss 
in section \ref{sec:nonmarkov} a non-Markovian master equation and 
show that for atoms the Born-approximation is not valid for our 
parameters even if we avoid making the Markov approximation.  In the 
regimes where these approximations fail, the output field becomes 
correlated with the system and hence cannot be traced over to provide 
a master equation.  Finally we discuss ways of proceeding when the 
Born-Markov master equation fails.  The most straightforward method is 
that used in section \ref{sec:Output} where exact equations for the 
whole system and reservoir are obtained.

\section{Exact solutions} \label{sec:Output}

Dilute gas Bose Einstein Condensates (BEC) are now available in the 
laboratory.  To produce a continuously running atom laser from a BEC 
requires the addition of a suitable pumping mechanism, and an output 
coupler.  A change of atomic state through RF transition has 
been used to produce a ``pulsed atom laser''\cite{MITExpt,MITExpt2}.  
The output coupling from a single mode to a large reservoir is 
sometimes described by a Born-Markov master equation of the form %
\begin{eqnarray}
\frac{d \rho(t)}{dt} &=& C (2 a \rho(t) a^{\dag} -  
a^{\dag} a \rho(t) -  \rho(t) a^{\dag} a),
\label{Eq.QuantumOpticalEqn}
\end{eqnarray}
where $C$ describes the strength of the coupling and $a$ ($a^{\dag}$) is 
the annihilation (creation) operator for the single mode system.  Two 
important approximations involved in such a description are that the 
evolution is Markovian and that it depends only on the system operators.  
The Markovian property means that the rate of change of the state depends 
only on the state at that time.  There is no explicit dependence on the 
state at previous times.  The equation is a function of system variables 
only, due to tracing over the reservoir.  This is appropriate if the 
reservoir remains uncorrelated with the system.  In fact, it is 
approximated in deriving Eq.  (\ref{Eq.QuantumOpticalEqn}) that the 
reservoir does not change with time.

We wish to investigate these two approximations, and their validity 
for describing atomic output coupling.  In a full atom laser model it 
is essential that a pumping term is also included.  However, in this 
paper we consider a single mode coupled only through output coupling 
to the outside world.  Experimentally this corresponds to a leaky BEC. 
We focus on the output coupling term which would be present in a full 
master equation model alongside other terms.

We begin by considering a generic output coupling mechanism which we 
have previously analysed in the context of Heisenberg equations of 
motion\cite{Hope97,Moy97b}.  Here we investigate how such a model can 
be described by a density operator equation - in particular a master 
equation.  When the output field is constrained by a waveguide such as 
in a hollow optical fibre model\cite{Moy97}, the output field is 
effectively one dimensional.  We consider a single mode system (the 
lasing mode with creation operators $a^{\dag}$) coupled to a 
one-dimensional continuum of free space modes.  The free space modes 
are labelled by their momentum, $\hbar k$ (creation operators 
$b_{k}^{\dag}$).  In a general three dimensional model $b_{k}$ would 
also be labelled by indicies describing the transverse modes.  Here, 
however, we suppress these indicies and assume that the output is 
coupled into a waveguide with the atoms remaining in a single 
transverse mode.  The Hamiltonian describing the system is given by %
\begin{eqnarray}
{ H} &=& { H_{\mbox{s}}} + 
      { H_{\mbox{r}}}  + { H_{\mbox{sr}}}, \label{Eq.Heff}\\
{ H_{\mbox{s}}} &=& \hbar \omega_{0} ~ a^{\dag} a, 
\label{Eq.Hsys}\\
{ H_{\mbox{r}}} &=& \int_{-\infty}^{\infty} \mbox{dk} ~ \hbar \omega_{k} ~ 
      b_{k}^{\dag} b_{k}, \label{Eq.Hext}\\
{ H_{\mbox{sr}}} &=& -i \hbar \int_{-\infty}^{\infty} \mbox{dk}~~ 
(\kappa (k) ~b_{k} a^{\dag} - \kappa^{*} (k)~ b_{k}^{\dag} a).  
\label{Eq.Hint}
\end{eqnarray} 
The function $\kappa(k)$ describes the shape of the coupling in $k$ 
space, and is left general here.  By appropriately choosing 
$\kappa(k)$ we may simulate a wide range of practical output coupling 
mechanisms, including position dependant coupling and the effect of 
momentum kicks.  The latter may result from laser photons when an atom 
undergoes a change of state.  Choosing $\kappa(k)$ as constant over a 
broad region corresponds to broadband coupling.  This model is used 
for optical input-output theory\cite{OpticalModel}, and in proposed 
atom laser theories which result in Born-Markov master equations.  For 
the atom case this model ignores potentially important effects such as 
gravity and atom-atom interactions.  For now, we neglect these effects 
in order to investigate the validity of the Born-Markov 
approximations.  We use this model to understand the differences 
between optical and atomic output coupling.  We will extend the model 
in section \ref{sec:nonzero}.

We use the Hamiltonian presented above to write Heisenberg equations 
of motion for the operators $a$, $b_{k}$.  We can also obtain 
equations for combinations of these operators which may be of more 
interest, such as the number of atoms in the system, $a^{\dag} a$.  
These equations can be difficult to solve in general.  However, since 
they include the output and system explicitly, they describe the 
dynamics of the model exactly\cite{Hope97,Moy97b}.  Exact solutions 
can be compared with Born-Markov master equations.

To facilitate this comparison we present exact equations for the 
expectation values of the system operators, $a^{\dag}$ and $a^{\dag} 
a$.  In these we assume (as we will do in the master equation 
descriptions to follow) that the reservoir is initially empty, 
$\langle b_{k}^{\dag}(0) b_{k}(0) \rangle = 0$.  We do not place any 
further restrictions on $\langle b_{k}^{\dag}(t) b_{k}(t) \rangle$.  
This is the first fundamental difference between the atom and photon 
case.  An empty reservoir for photons is inapropriate at finite 
temperatures\cite{Carmichael91}.  In experiments where a Bose-Einstein 
condensate is allowed to leak out of a trap, however, the most 
appropriate initial state for the outside atom modes is a vacuum.  
Similarly, for an atom laser in a hollow fibre, which we discussed in 
\cite{Moy97}, the initial output modes would be empty.

We obtain %
\begin{eqnarray}
\frac{d}{d t} \langle a^{\dag} (t)\rangle &=& i \omega_{0} \langle 
a^{\dag} (t) \rangle - \nonumber\\
&&~  \int_{0}^{t} d \tau ~ f'^{*}(\tau)  \langle 
a^{\dag}(t-\tau) \rangle e^{i \omega_{0} \tau}, \label{Eq.adag} \\
\frac{d}{d t} \langle a^{\dag}(t) a(t) \rangle &=& - \int_{0}^{t} d 
\tau ~ f'(\tau) ~ \langle a^{\dag}(t) a(t-\tau) \rangle e^{-i 
\omega_{0} \tau} ~ \nonumber \\
  && ~+ \mbox{h.c}. \label{Eq.adaga}
\end{eqnarray}
These equations have solutions given by \cite{Hope97,Moy97b}
\begin{eqnarray}
\langle a^{\dag} (t)\rangle  &=& \langle a^{\dag}(0) \rangle e^{i 
\omega_{0} t} ~{\cal 
L}^{-1} \left \{ \frac{1}{s + {\cal L} \left\{f'(t)^{*} \right\}(s) } 
\right\} (t), \label{Eq.adagsoln}
\end{eqnarray}
\begin{eqnarray}
\langle a^{\dag}(t) a (t)\rangle &=& \langle a^{\dag}(0) a(0) \rangle 
~ \times \nonumber\\
&& ~ \left| {\cal L}^{-1} \left \{ \frac{1}{s + {\cal L} \left\{f'(t) 
\right\}(s) } \right\} (t) \right|^{2}, \label{Eq.adagasoln}
\end{eqnarray}
where ${\cal L}$ and ${\cal L}^{-1}$ denote Laplace and inverse 
Laplace transforms respectively.  $f'(t)$ is the function defined by 
\cite{Moy97b} %
\begin{eqnarray}
f'(t) &=& \int_{-\infty}^{\infty} dk |\kappa(k)|^{2} e^{-i 
(\omega_{k}-\omega_{0}) t}. 
\label{Eq.f't}
\end{eqnarray}
The function $f(t) = f'(t) e^{-i \omega_{0} t}$ is the ``reservoir 
correlation function'' in the master equation picture which we 
describe in the next section.  Here $\omega_{k} = \hbar k^{2}/(2 m)$ 
for atoms, in contrast to $\omega_{k} = c_{L} |k|$ for photons.  Here 
$m$ is the mass of the atoms, and $c_{L}$ is the speed of light.  

For a previously described output coupling from a condensate through state 
change\cite{Moy97b}, $\kappa(k)$ is a Gaussian of width $\sigma_{k}$, %
\begin{eqnarray}
\kappa(k) &=& i \Gamma^{\frac{1}{2}} (2 \pi \sigma_{k}^{2})^{-1/4} 
\exp{\left(-k^{2} / (4 \sigma_{k}^{2}) \right) }. \label{Eq.kappa}
\end{eqnarray}
Here, the strength of the coupling is given by the coupling constant, 
$\Gamma$.  For this form of coupling $f'(t)$ may be evaluated as: %
\begin{equation}
f'(\tau) = \frac{e^{i \omega_{0} \tau} \Gamma}{\sqrt{1 + i \alpha \tau}}, 
\label{Eq.f'b}
\end{equation}
where we have defined
\begin{equation}
\alpha = \hbar \sigma_{k}^{2}/m. \label{Eq.alpha}
\end{equation}

For the broadband limit of Eq.~(\ref{Eq.alpha}) (discussed in section 
\ref{sec:Validity}) we may use methods similar 
to those given in reference \cite{Moy97b} to obtain an analytical form for 
the inverse Laplace transform,
\begin{eqnarray}
{\cal L}^{-1} \left \{ \frac{1}{s + {\cal L} \left\{f'(t) 
\right\}(s) } \right\} (t)  &=& e^{i \omega_{0} t} \nonumber\\
    &&\rule{-35mm}{0cm}\times ~(W(a,b,c) + 
  W(b,a,c) +  W(c,b,a) ),  \label{Eq.LaplaceSoln} \\
W(a,b,c) &=& \frac{a^{2} e^{a^{2} t} ~ (1 + \mbox{erf}[a 
\sqrt{t}])}{(a-b)(a-c)}. \label{Eq.Wabc} 
\end{eqnarray}
The variables, a,b and c are the three solutions of the equation $s^{3} + i
\omega_{0} s + \Gamma c \sqrt{i} = 0$.  Note that Eq.~(\ref{Eq.adagsoln}) 
gives that $\langle a(t) \rangle$ will always remain zero if $\langle a(0) 
\rangle = 0$.  For the case of damping of a BEC which we consider here, the 
initial state corresponds to a BEC in an atom trap.  According to 
spontaneous symmetry breaking arguments BECs are in coherent states with a 
definite global phase \cite{StatMech}, so that $\langle a(0) \rangle \neq 
0$.  This is a useful assumption.  Nevertheless, even if $\langle a \rangle 
= 0$, the form of the equation for $\langle a(t)
\rangle$ must be as given.

The equations of motion given by Eqs.  (\ref{Eq.adag},\ref{Eq.adaga}) 
and the corresponding solutions, Eqs.  
(\ref{Eq.adagsoln},\ref{Eq.adagasoln}) are exact for the system under 
consideration.  In specific cases it is very difficult to solve for 
these inverse Laplace transforms.  Moreover, the Heisenberg equations 
for system operators depend on the external operators, $b_{k}(0)$ in 
general.  We next investigate equations of motion based on the Born 
and Markov approximations.  These are compared with the exact 
solutions given above.

\section{The Born-Markov Master Equation} \label{sec:MasterEqn}

Derivations of the Born-Markov master equation for a general system 
reservoir interaction are given in references
\cite{Carmichael91,Louisell90,Tannoudjii92}.  Here we present 
a derivation for the specific model of Eq.  (\ref{Eq.Heff}) to 
Eq.(\ref{Eq.Hint}).  We assume that the atom reservoir is initially in 
a vacuum state - that is, there are initially no atoms outside the 
system.  This assumption was also made in the exact solutions 
presented in Sec.  \ref{sec:Output}.  We make the Born approximation.  
It involves ignoring correlations which may arise between the system 
and reservoir and ignoring any time evolution of the reservoir density 
operator.  We use the interaction Hamiltonian in Eq.  (\ref{Eq.Hint}).  
This leads to the non-Markovian master equation: %
\begin{eqnarray}
\frac{d{\widetilde{\rho}}}{dt} &=& - \int_{0}^{t}~ d\tau \left\{ a^{\dag}a 
\widetilde{\rho}(t-\tau) - a \widetilde{\rho}(t-\tau) a^{\dag} 
\right\} \times \nonumber\\
&& f'(\tau) ~ + \mbox{h.c}, \label{Eq.Born} 
\end{eqnarray}
where, ${\widetilde{\rho}}$ is the density operator in the interaction 
picture.  The function, $f'(\tau)$, defined as $f'(\tau) = f(\tau)~ e^{i 
\omega_{0} \tau}$ is the same as defined in Eq.  (\ref{Eq.f't}) and 
Eq.~(\ref{Eq.f'b}).  Here $f(\tau)$ is the reservoir correlation function.

Eq.  (\ref{Eq.Born}) is non-Markovian.  The second major approximation 
required to produce a Born-Markov master equation is the Markov 
approximation.  The Markov approximation is made on the assumption that the 
reservoir correlation function, $f(\tau)$ goes to zero rapidly compared 
with the time scale on which $\widetilde{\rho}(t)$ changes.  Making the 
Markov approximation thus involves replacing the terms $\widetilde{\rho}(t-
\tau)$ in Eq.  (\ref{Eq.Born}) with $\widetilde{\rho}(t)$.  In the optics 
case, this approximation is also usually accompanied with the extension of 
the upper limit of the integral from $t$ to infinity.  Making these 
approximations gives %
\begin{eqnarray}
\dot{\widetilde{\rho}} &=& \left\{ (c + c^{*}) a \widetilde{\rho}(t) 
a^{\dag} - c~ a^{\dag}a \widetilde{\rho}(t) - c^{*} 
\widetilde{\rho}(t) a^{\dag} a \right\} , \label{Eq.Markov}
\end{eqnarray}
where
\begin{eqnarray}
c &=&  \int_{0}^{t \rightarrow \infty} d \tau 
\int_{-\infty}^{\infty} |\kappa(k)|^{2} e^{-i (\omega_{k}-\omega_{0}) 
\tau} dk \nonumber\\
&=& \int_{0}^{\infty} f'(\tau) d\tau \nonumber\\
 &=& \frac{\Gamma \sqrt{2 \pi} \exp{\left[- \omega_{0} / \alpha
\right]}}{\sqrt{2 \omega_{0} \alpha }} ~ \left(1 + \mbox{Erf} \left[ i 
\sqrt{\frac{\omega_{0}}{ \alpha}} \right] \right). \label{Eq.c}
\end{eqnarray}

The upper integration limit $t$ has been extended to $\infty$, as is 
done in the optical case, without affecting our subsequent conclusion 
regarding the Born and Markov approximations.  This produces an 
equation of the form which has been used previously to describe atom 
lasers.  We further note that we could redefine $c$ to be real without 
loss of generality by incorporating the imaginary part of $c$ in with 
the free part of the system Hamiltonian.  This reduces the form of the 
loss term to the familiar $C (2 a \rho a^{\dag} - a^{\dag}a \rho - 
\rho a^{\dag} a) $, with $C = {\cal R}e[c]$ the (real) coupling 
strength.

The value of the constant $c$ depends on the form of $f'(\tau)$.  In the 
following, we consider $f'(t)$ to be either that resulting from the 
Gaussian coupling presented in this section, or the broadband limit of this 
function which we discuss in the next section.

\section{Timescale conditions} \label{sec:Validity}

The Born-Markov master equation, Eq.  (\ref{Eq.Markov}), is the form 
used recently in various discussions of atom laser dynamics 
\cite{Holland95,Wiseman95a,Wiseman95b,Guzman96,Wiseman97}.  This 
Born-Markov master equation is used ubiquitously in quantum optics.  
The validity of the Markov approximation depends on the interplay 
between the reservoir correlation timescale, the system timescale, and 
the timescale on which the system decays.

The condition for the validity of the Born-Markov approximations is 
given in standard optics texts by the timescale separation condition 
\cite{Carmichael91,Louisell90,Tannoudjii92}
\begin{equation}
t_{R} << \Delta t << t_{D}
\end{equation}
where $t_{R}$ is the reservoir correlation time, and $t_{D}$ is the 
cavity decay time.  $\Delta t$ defines a coarse grained timescale on 
which the equations of motion are valid.  Generally $t_{R}$ is defined 
as the ``timescale on which the reservoir correlations are non-zero''.  
$t_{D}$ is the decay timescale of the system which can be obtained by 
solving the equations for system and reservoir self consistently.  In 
the Markov approximation this timescale goes as $1/(c + c^{*})$ for 
$c$ similar to that defined in Eq.  (\ref{Eq.c}) except with the 
optics dispersion relation.

Both $t_{R}$ and $t_{D}$ depend on the function $f'(\tau)$, and thus 
in turn on the form of $\omega_{k}$ as a function of $k$.  For atoms 
$\omega_{k} = \hbar k^{2}/(2 m)$, whereas for photons $\omega_{k} = 
c_{L} |k|$, where $c_{L}$ is the speed of light.  It also depends on 
factors such as the nature of the reservoir and the parameter regime 
in which atom lasers operate. 

A second timescale condition for the Born-Markov approximation which 
is discussed in some treatements \cite{Louisell90,Tannoudjii92} of the 
optical Born-Markov approximation is that the system timescale, 
defined as $t_{s} = 1/\omega_{0}$ must satisfy
\begin{equation}
t_{s} << \Delta t. \label{Eq.condition2}
\end{equation}
This is equivalent to requiring $\omega_{0}$ to be very large.  A large 
$\omega_{0}$ condition is required in optics for a number of reasons.  
First, the initial coupling Hamiltonian, of the form ($a b_{k}^{\dag} + 
a^{\dag} b_{k})$, is in the rotating wave approximation and ignores terms 
of the form $a b_{k}$ and $a^{\dag} b_{k}^{\dag}$.  This rotating wave 
approximation in optics can only be made in the case of large $\omega_{0}$.  
For the atom coupling, however, the correct form of the coupling does not 
include such (non atom number conserving) terms, even for small 
$\omega_{0}$.  The terms which are eliminated in the optics 
case\cite{Louisell90,Tannoudjii92} through the assumption of large 
$\omega_{0}$ are already zero for our model, due to the assumption of an 
atom vacuum reservoir.  Thus, one may be led from these treatments to 
suppose that the Born-Markov approximation is made independently of 
$\omega_{0}$ for an atom-vacuum reservoir.  This however is not the case as 
we will discuss later.

In optics, these timescale conditions are usually satisfied.  For a 
coupling based on a mirror it is standard to assume that the coupling 
is broadband\cite{QuantumOptics}.  That is, we assume $\kappa(k)$ is 
flat in k-space.  In this case the reservoir correlation function, 
$f'(\tau)$, given by Eq.  (\ref{Eq.f't}) becomes %
\begin{eqnarray}
f'(\tau) &\approx& |\kappa(k_{0})|^{2} e^{i \omega_{0}\tau} 
~\int_{-\infty}^{\infty} ~ e^{-i \omega_{k} \tau}~ dk \nonumber\\
 &=& 2 |\kappa(k_{0})|^{2}  
~\int_{-\omega_{0}/c_{L}}^{\infty} ~ e^{-i c_{L} k \tau}~ dk \nonumber\\
&\approx& 2 |\kappa(k_{0})|^{2} 
~ \delta(\tau). \label{Eq.opticcompare1}
\end{eqnarray}
In the final equation the Dirac delta function, $\delta(\tau)$, is 
obtained by extending the frequency integral into physically 
unrealistic negative frequencies.  This is a standard technique in 
optics \cite{QuantumOptics} where $\omega_{0}$ is typically large.

Typically, for a laser, $\omega_{0} \approx 10^{15} \mbox{s}^{-1}$ is 
large and the assumption of a Dirac delta function decay is very good.  
Atom traps work in rather different parameter regimes with $\omega_{0} 
\approx 10^{3} \mbox{s}^{-1}$.  If we avoid using negative 
frequencies, with the assumption of the empty reservoir considered 
here, we obtain the sum of a Dirac delta function, and an imaginary 
part corresponding to the Cauchy principal value of the integral in 
Eq.  (\ref{Eq.opticcompare1}).
\begin{equation}
f'(\tau) = \frac{2 \pi}{c_{L}} |\kappa(k_{0})|^{2} e^{i \omega_{0} 
\tau} \left( \delta(\tau) - \frac{i}{\pi \tau} \right).
\end{equation}
However, for an optical reservoir this estimate of correlation function 
decay based on our reservoir model is not strictly valid.  This is because, 
while we have considered here the photonic dispersion relation, a vacuum is 
unrealistic for an optical reservoir at finite temperatures.  More 
appropriate is a thermal reservoir, which leads to a decay time of order 
$\hbar / k_{B} T
\approx 10^{-13} \mbox{s}$\cite{Carmichael91}.

These reservoir correlation times must be short compared with the 
decay time, $t_{D}$.  The system timescale, $1/\omega_{0}$ must also 
be short compared with $t_{D}$ for a standard optical reservoir.  
$t_{D}$ is the timescale of exponential decay, given by $e^{-(c+c^{*}) 
t}$ in the Born-Markov limit.  From an equivalent derivation of the 
optical master equation to that given in section \ref{sec:MasterEqn}, 
the decay timescale is given by $t_{D} = 1/(c+c^{*}) \propto 
(\sigma_{k}/\Gamma)$.  That is $t_{D}$ is inversely proportional to 
the strength of the coupling, given by $\Gamma/\sigma_{k}$ in the 
broadband limit.  We will see later that for the atom coupling, the 
different dispersion relation makes $t_{D}$ depend on $\omega_{0}$ and 
other parameters, such as the mass of the atom.  For optical systems 
this decay time is typically of the order $10^{-8} \mbox{s}$.  Thus, 
in typical optical systems, the Born-Markov approximation holds for a 
number of reasons.  The condition $t_{R} << t_{D}$ holds, because in 
the large $\omega_{0}$ limit the reservoir correlations decay as a 
Dirac delta function.  $t_{D}$ does not depend on $\omega_{0}$ and is 
typically many orders of magnitude larger than the reservoir decay 
times.  Similarly, the system timescale $t_{s}$ for realistic optical 
cavities is very much shorter than the decay timescale, $t_{D}$.

In contrast, the large $\omega_{0}$ limit is not necessarily valid for 
realistic atom traps.  Moreover, even in the limit of infinitely large 
$\omega_{0}$, the atom correlation function $f'(\tau)$ does not tend 
towards a Dirac delta function.  This is due to the atomic dispersion 
relations, which lead to a dependence of $t_{D}$ on parameters other 
than the coupling strength.  Furthermore, the assumption of an 
initially empty reservoir is realistic for atoms.  For atoms, the 
broadband limit of the function $f'(t)$ is given by %
\begin{eqnarray}
f'(\tau) &\approx& |\kappa(k_{0})|^{2} e^{i \omega_{0}\tau} 
~\int_{-\infty}^{\infty} ~ e^{-i \omega_{k} \tau}~ dk \nonumber\\
&=& 2 |\kappa(k_{0})|^{2} e^{i \omega_{0}\tau} 
~\int_{0}^{\infty} ~ e^{-i \hbar k^{2}/(2 m) \tau}~ dk 
\nonumber\\
 &=&  |\kappa(k_{0})|^{2} ~e^{i \omega_{0}\tau} ~\frac{\sqrt{m 
 \pi} \left(1 - i  \right)}{\sqrt{\hbar \tau}}.  
 \label{Eq.opticcompare2}
\end{eqnarray}
This is the broadband limit of the general reservoir 
correlation function, Eq.  (\ref{Eq.f'b}) given in Sec.  
\ref{sec:MasterEqn}.  Both broadband and Gaussian coupling give forms for 
$f'(\tau)$ which fall off as inverse $\sqrt{\tau}$.  The broadband limit of 
Eq.  (\ref{Eq.f'b}) is obtained as both $\sigma_{k}$ and $\Gamma$ tend to 
infinity, with $\Gamma/\sigma_{k} =
\mbox{const.} = \sqrt{2 \pi} |\kappa(k_{0})|^{2}$.  $\sigma_{k}$ and 
$\Gamma$ are both defined in Sec.  \ref{sec:MasterEqn} with 
$\sigma_{k}$ giving the width of the Gaussian coupling, $\kappa(k)$.  
We note that the broadband limit of Eq.  (\ref{Eq.f'b}) is 
not correctly obtained by taking $\sigma_{k} \rightarrow \infty$ while 
keeping $\Gamma$ constant.  If $\sigma_{k} \rightarrow \infty$ then 
the coupling function $\kappa (k)$ and the constant $c$ in the master 
equation, Eq.  (\ref{Eq.Markov}), go to zero everywhere due to the 
normalisation of our coupling, Eq.~(\ref{Eq.kappa}).  

We may now highlight three significant differences between optical and 
atomic output coupling.  Firstly, Eq.~(\ref{Eq.opticcompare2}) shows that 
the atomic reservoir correlation function decays as $1/\sqrt{\tau}$.  This 
is different from the optical case even for finite $\omega_{0}$, and will 
lead to different behaviour of the exact equations.  Secondly, the system 
timescale, $1/\omega_{0}$, is significantly larger for atom traps than for 
optical cavities.  Thirdly, the timescale on which the system decays is 
given by $t_{D} \approx 1 / (c + c^{*})$ where $c$ is related to the 
integral of the correlation function, $f'$, given in Eq.~(\ref{Eq.c}).

The optical dispersion relation causes the system decay time to depend 
only on the coupling strength, $\Gamma/\sigma_{k}$, and the speed of 
light, $c_{L}$.  For the atom dispersion relation, $t_{D}$ also 
depends on the system frequency, $\omega_{0}$ and the mass of the 
atom, $m$, and is given by
\begin{equation}
t_{D} = 2 \sqrt{\frac{\omega_{0} \hbar}{m \pi}} 
\left(\frac{\sigma_{k}}{\Gamma} \right) \label{Eq.tD}
\end{equation}

These timescale considerations based on the differing dispersion relations, 
reservoir model, and parameter regimes for atoms compared with optics 
suggest that the Born-Markov approximation may not be generally valid in 
describing output from practical atom lasers.  In fact, the optical 
Born-Markov theory is not universally valid for optics, either.  In a 
photonic band gap material the dispersion relation for the photons and the 
radiation reservoir may be modified.  Bay {\it et al.} \cite{Bay97} have 
discussed fluorescence into a radiation continuum in which a band gap with 
dispersion relation near the band edge, $\omega_{k} = \omega_{e} + A 
(k-k_{0})^{2}$, similar to the atomic dispersion relation, is present.  
They find behaviour which cannot be described using the Born-Markov theory.

In the next section, we discuss equations of motion obtained from the 
Born-Markov master equation for $\langle a^{\dag}(t) \rangle$ and 
$\langle a^{\dag}(t) a (t) \rangle$.  We will compare these with those 
obtained from the exact equations, Eq.  (\ref{Eq.adag}) and Eq.  
(\ref{Eq.adaga}) and hence examine the regimes of validity of the 
Born-Markov approximation.

\section{The validity of the Born-Markov approximation.}
\label{sec:equations}

We consider first the Born-Markov master equation, Eq. 
(\ref{Eq.Markov}).  Using the relation,  
\begin{equation}
\frac{d \langle \tilde{o} \rangle}{dt} = \mbox{Tr} \left[\tilde{o} 
\frac{d \tilde{\rho}}{d t} 
\right], \nonumber
\end{equation}
where $\tilde{o}$ is any system operator in the interaction picture, we 
obtain the following equations of motion for $\langle a^{\dag} \rangle$ and 
$\langle a^{\dag} a \rangle$ from the Markovian master equation, Eq.  
(\ref{Eq.Markov}).
\begin{eqnarray}
\frac{d \langle a^{\dag}(t) \rangle}{dt} &=& \left[i \omega_{0} - 
c^{*} \right] \langle a^{\dag} \rangle (t),\label{Eq.dadagdt}\\
\frac{d \langle a^{\dag}(t) a(t) \rangle}{dt} &=& -(c + c^{*}) 
\langle a^{\dag}(t) a(t) \rangle. \label{Eq.dadagadt}
\end{eqnarray}

We compare the exact equations, Eq.  (\ref{Eq.adag}) and Eq.  
(\ref{Eq.adaga}) with Eq.  (\ref{Eq.dadagdt}) and Eq.  
(\ref{Eq.dadagadt}) respectively.  The equations derived from the 
Born-Markov master equations are equivalent to the exact equations 
under the approximation that the term $\langle a^{\dag}(t-\tau) 
\rangle e^{i \omega_{0} \tau} = \langle a^{\dag}(t) \rangle$ and 
$\langle a^{\dag}(t) a(t-\tau) \rangle e^{-i \omega_{0} \tau} = 
\langle a^{\dag}(t) a(t) \rangle$.  That is, if we ignore the effect 
of the interaction on the system evolution.  Alternatively, the exact 
and Born-Markov equations will agree if $f'(\tau)$ can be approximated 
by a Dirac delta function.  For atoms, however, there is no limit in 
which $f'(\tau)$ tends to a Dirac delta function.

When $f'(\tau)$ is not given by a Dirac delta function, the 
approximation to replace $\langle a^{\dag}(t-\tau) \rangle e^{i 
\omega_{0} \tau}$ by $\langle a^{\dag}(t) \rangle$ in Eq.  
(\ref{Eq.adag}) will still be valid in some parameter regimes.  In 
particular, if we consider the exact equation, Eq. (\ref{Eq.adaga}), 
and the solutions obtained from the Heisenberg equations of motion, we 
can see that the exact equation can be rewritten as
\begin{eqnarray}
\frac{d}{d t} \langle a^{\dag}(t) a(t) \rangle = - \langle 
a^{\dag}(t) a(t) \rangle \times \int_{0}^{t} ~ f'(\tau) ~ \nonumber \\
 \times \frac{{\cal L}^{-1} \left \{ \frac{1}{s 
 + {\cal L} \left\{f'(t) \right\}(s) } \right\} (t-\tau)}{{\cal 
 L}^{-1} \left \{ \frac{1}{s + {\cal L} \left\{f'(t) \right\}(s) } 
 \right\} (t)} ~d \tau ~+ \mbox{h.c}.  \label{Eq.newadaga}
\end{eqnarray}
From this form of the exact equation, the Born-Markov equation is 
obtained by the assumption that $f'(\tau)$ decays rapidly on the 
timescale on which the other (inverse Laplace transform) terms in the 
integral change with $\tau$.  For parameters in which the Born-Markov 
approximation is valid, we know that this ratio, given exactly from 
Eq.  (\ref{Eq.LaplaceSoln}), is approximately exponential with a 
timescale of order $t_{D}$.  This fact can be motivated by considering 
Eq~(\ref{Eq.adagasoln}).  This equation shows that the number of atoms 
in the cavity as a function of time is given by the square of the 
absolute value of the inverse Laplace transform term, identical to 
that found in Eq.~(\ref{Eq.newadaga}) above.  In the Born-Markov 
regime we know the number of atoms in the cavity must decay 
approximately exponentially on the timescale $t_{D}$.  Thus, for the 
Born-Markov approximation to hold we require that the the timescale on 
which $f'(t)$ decays must be small compared to $t_{D}$.

For the non-broadband case, we can define a timescale on which $f'(t)$ 
decays by the width at half maximum of the absolute value of the 
reservoir correlation function $|f(\tau)|$ %
\begin{eqnarray}
t_{R} &=& \frac{\sqrt{15} m}{\hbar \sigma_{k}^{2}} = 
\frac{\sqrt{15}}{\alpha}. \label{Eq.tr}
\end{eqnarray}
Here $m$ is the mass of the atoms and $\sigma_{k}$ is the width of the 
Gaussian lasing mode in momentum space.  This timescale must be small 
compared with the decay timescale, $t_{D}$ discussed earlier.  This 
condition, by itself, would suggest that as the coupling becomes 
increasingly broadband the Born-Markov approximations become increasingly 
good.  However, this is not the case.  Although the function $f'(\tau)$ 
becomes infinite at zero in the broadband limit and therefore must have a 
zero half width, $t_{R}$, there are significant contributions to the 
integral in Eq.~(\ref{Eq.newadaga}) away from $\tau=0$.  Instead of the 
half width of the reservoir correlation function, $t_{R}$, we are more 
interested in the half width of the integral of $f'(\tau)$.  This timescale 
is defined in terms of the reservoir correlation function, $f(\tau)$, and 
the system frequency, $\omega_{0}$.

We define $t_{rs}$ such that 
\begin{equation}
	\int_{0}^{t_{rs}} f'(\tau) d\tau = 1/2 
	\int_{0}^{\infty} f'(\tau) d\tau.
	\label{Eq.trs}
\end{equation}

For broadband coupling we find that $t_{rs} = 1/\omega_{0}$ is 
equivalent to the system timescale, $t_{s}$, defined earlier.  The 
atomic dispersion relations make $t_{D}$ also depend on $\omega_{0}$, 
Eq.~(\ref{Eq.tD}) which means that the condition $t_{rs} << t_{D}$, can 
be written as
\begin{eqnarray}
\omega_{0}^{3/2} \sqrt{\frac{\hbar}{m \pi}} 
\left(\frac{\sigma_{k}}{\Gamma} \right) >> 1. \label{Eq.inequality}
\end{eqnarray}

For broadband coupling, this timescale condition determines the 
parameter regimes in which the Born-Markov approximation is valid.  
The dependence on $\omega_{0}^{3/2}$ and on the mass in this condition 
comes from the dependence of $t_{D}$ on $\mbox{m}$ and $\omega_{0}$.  
In the equivalent model with optical dispersion relations, $t_{D}$ 
only depends on the strength of the coupling, given by 
$\Gamma/\sigma_{k}$.  

We now compare the results of the exact and Born-Markov equation.  We 
initially consider Gaussian coupling with similar parameters to those 
discussed in \cite{Moy97b}.  For atoms, $m \approx 5 \times 10^{-26} 
\mbox{kg}$.  Atom traps in which BEC has been achieved have 
frequencies of $\omega_{0} \approx 2 \pi \times 123 
\mbox{s}^{-1}$\cite{MIT3}.  Values for the coupling strength, $\Gamma$ 
depend on the method used.  For Raman coupling, for instance, $\Gamma$ 
depends on the intensity of the lasers\cite{Moy97b}, so a range of 
values down to zero can be achieved.  The value $\Gamma \approx 10^{6} 
\mbox{s}^{-2}$ can be achieved with laser intensities similar to those 
discussed in \cite{Moy97b}.  The width of the Gaussian we assume to be 
determined by $\sigma_{k} \approx 10^{6} \mbox{m}^{-1}$ corresponding 
to a ground state wavefunction of size $\approx 2 \mu \mbox{m}$.

\begin{figure}
\begin{center}
\epsfxsize=\columnwidth
\epsfbox{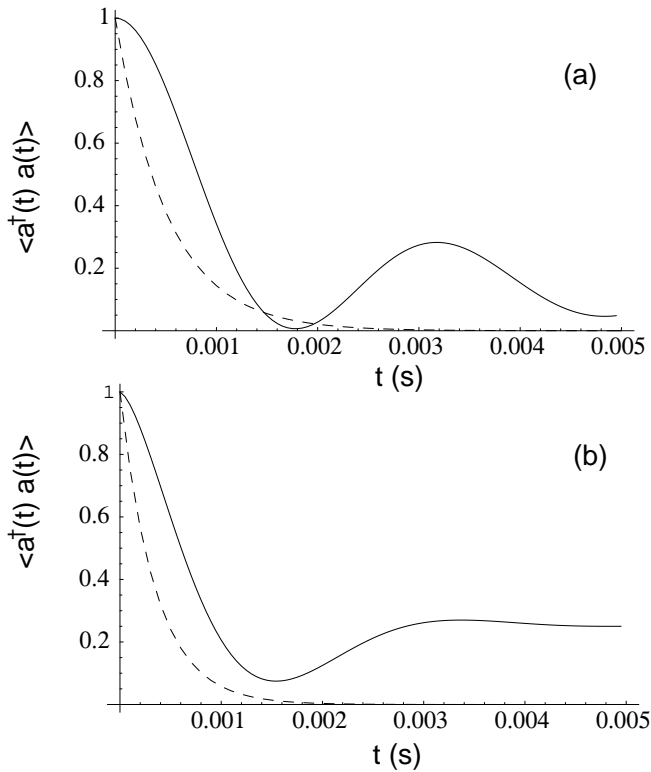}
\end{center}
\caption{A comparison between $\langle a^{\dag}(t) a(t) \rangle $ 
found using the Born-Markov master equation (dashed line) and the 
exact solution (solid line).  Atom numbers are normalized so that 
$\langle a^{\dag}(0) a(0) \rangle = 1$.  Parameters are (a) $\Gamma = 
1 \times 10^{6} \mbox{s}^{-2}$, $\sigma_{k} = 10^{6} \mbox{m}^{-1}$ 
for Gaussian coupling and (b) $|\kappa(k_{0})|^{2} = 1.0/\sqrt{2 \pi} 
\mbox{m}\mbox{s}^{-2}$ for broadband coupling.  Other parameters are 
$m = 5 \times 10^{-26} \mbox{kg}$, $\omega_{0} = 2 \pi \times 123~ 
\mbox{s}^{-1}$.}
\label{Fig.1}
\end{figure}

In Fig \ref{Fig.1}a the solution for the expectation value of the 
number of atoms in the system at time $t$ is plotted for the 
parameters quoted above.  The exact solution is given by Eq.  
(\ref{Eq.adagasoln}).  The solution derived from the Born-Markov 
master equation is also shown.  This is given by %
\begin{eqnarray}
\langle a^{\dag}(t) a(t) \rangle &=& \langle a^{\dag}(0) a(0) \rangle 
~ e^{ -(c+c^{*})t}.  \label{Eq.dadagasolnbm}
\end{eqnarray}
Figure \ref{Fig.1} demonstrates that the results for the number of 
atoms using the Born-Markov approximations disagree with the exact 
solutions in the case of Gaussian output coupling.  For these 
parameters, $t_{R} \approx 2.0 \times 10^{-3} \mbox{s}$, $t_{s} 
\approx 1.4 \times 10^{-3} \mbox{s}$ and $t_{D} \approx 5.0 \times 
10^{-4} \mbox{s}$,  so that both the inequalities $t_{R} < t_{D}$ 
and $t_{s} < t_{D}$ fail.

We have presented results in terms of the Gaussian coupling discussed in 
Sec.  (\ref{sec:MasterEqn}).  This reflects a possible realistic output 
coupling method.  However from the inequality $t_{s} < t_{D}$ we see that 
even in the broadband limit, atoms do not couple out of the system in a 
Born-Markov manner for parameters which correspond to the ones we discuss 
above.  Figure \ref{Fig.1}b compares the exact and Born-Markov solutions to 
our model in the broadband limit.  In this limit the timescale, $t_{R}$ as 
defined above tends to zero, thus the first inequality is satisfied.  
However, the inequality, Eq.  (\ref{Eq.inequality}), fails.  This figure 
demonstrates that even for infinitely broad atom coupling the Born-Markov 
approximation may fail.  The parameters chosen are the same as those for 
the Gaussian coupling, with a strength of coupling in the exact solutions 
given by
\begin{equation}
|\kappa(k_{0})|^{2} = \frac{\Gamma}{\sqrt{2 \pi}~\sigma_{k}} \approx 
0.4 \mbox{m s}^{-2},
\label{Eq.kappa0}
\end{equation}
In this case the Born-Markov solution again gives exponential decay.  
However, the decay constant is now given as the broadband limit of Eq.  
(\ref{Eq.c}).

Figure \ref{Fig.1} demonstrates that for our model of output coupling 
from an atom laser, results for the number of atoms using the 
Born-Markov approximations disagree qualitatively with the exact 
solutions.  One of the effects of not being able to ignore the 
back-action from the reservoir is that for the model we are 
considering the number of atoms in the cavity does not tend to zero 
for long times (Figure \ref{Fig.1}).  We discuss reasons for this 
behaviour in the next section. If effects such as gravity and repulsive 
interactions are included the cavity number will tend to zero.  
However, even with these effects included the loss is not exponential 
and the conclusion that the Born-Markov approximation fails to 
describe the output coupling remains.

Here, we have demonstrated the failure of the Born-Markov 
approximation by the use of the particular system variable, $\langle 
a^{\dag} a \rangle$.  However, the problems with using the Born-Markov 
approximations are not confined to this particular example.  For 
instance, if the output from a BEC is described using a Born-Markov 
master equation, the resulting long time energy spectrum is 
Lorentzian.  However, if we avoid making the Born-Markov 
approximations for atoms the exact spectrum may be non-Lorentzian for 
some values of coupling strength, $\Gamma$, and frequency, 
$\omega_{0}$ \cite{Moy97b}.

\section{Effect of gravity on the model} \label{sec:nonzero}

In this section we present a quasi-single particle model which allows 
us to consider the effects of gravity on our output coupling.  Such a 
model cannot be extended to show interesting effects, such as noise 
suppression due to gain saturation if pumping is added to the model as 
it does not give information about the general quantum statistics of 
the output atom field.  However, it shows that the inclusion of 
gravity causes the atom number to asymptote to zero.  It does this in 
a non-exponential way, however, and therefore cannot be modelled by a 
damping term based on the Born-Markov approximation.  Moreover, we 
demonstrate that the short time behaviour predicted by the models with 
gravity agree with the exact solutions we have presented earlier which 
ignore the effects of gravity.

The previous section demonstrated qualitative differences between the 
exact solution of our model and the solution which uses the 
Born-Markov approximation.  In fact, the exact solution of the model 
has a stable, nondispersing state which means that not all of the 
atoms leave the cavity, whereas the approximate solution shows an 
exponential decay of atoms from the cavity.  The presence of the 
stable state is due to a coherence between the atoms in the cavity and 
the output modes.  The Born-Markov approximation ignores any coherence 
between the cavity mode and the output modes, and therefore cannot 
describe any model which produces such features.  We now show that 
such features would be destroyed by gravity.

For coherent dynamics without atom-atom interactions, the 
multiparticle evolution is identical to the evolution of a single 
particle \cite{Kasevich}.  The gravitational term makes it impossible 
to derive analytical results as in section ~\ref{sec:Output}, so we 
solved the corresponding time-dependent Schr\"{o}dinger equation 
numerically.  This was done in the position basis rather than the 
momentum basis for convenience.  The Hamiltonian for our system with 
the inclusion of a gravitational potential is
\begin{eqnarray}
H &=& H_{s} + H_{r} + H_{sr} + H_{g}, \label{Eq.H2} \\
H_{s} &=& \hbar \omega_{0} |\psi_{a} \rangle \langle \psi_{a} | 
,\label{Eq.Hs2} \\
H_{r} &=& \frac{\hat{P}^{2}}{2 M}, \label{Eq.Hr2} \\
H_{sr} &=& \int ~ dx~ \left( g^{*}(x) |\psi_{a} \rangle \langle x | ~+~ 
g(x) |x \rangle \langle \psi_{a} | \right), \label{Eq.Hsr2} \\
H_{g} &=& M g \hat{x} \sin(\theta), \label{Eq.Hgrav}
\end{eqnarray}
where the coupling function, $g(x)$ is related to $\kappa(k)$ by a 
Fourier transform,
\begin{eqnarray}
g(x) &=& \int~ dk \frac{i \hbar \kappa^{*}(k) ~ e^{i k x}}{\sqrt{2 
\pi}}. \label{Eq.gx}
\end{eqnarray}

\begin{figure}
\begin{center}
\epsfxsize=\columnwidth
\epsfbox{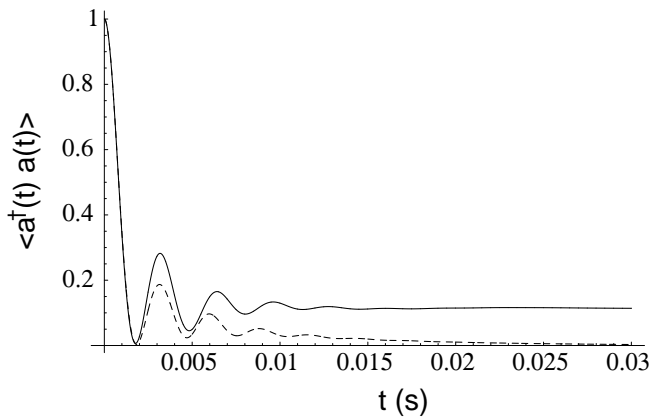}
\end{center}
\caption{The effect of introducing gravity into the model on the 
number of atoms in the cavity $\langle a^{\dag}(t) a(t) \rangle$.  
Atom numbers are normalized so that $\langle a^{\dag}(0) a(0) \rangle 
= 1$.  The solid line shows the figure with gravity turned off and 
agrees with the results presented in Fig.  \ref{Fig.1}a.  The dashed 
line includes gravity, $g ~\mbox{sin} (\theta)$, $g = 9.8 \mbox{m 
s}^{-1}, \theta = \pi/20$.}
\label{Fig.2}
\end{figure}

$\hat{P}$ is the momentum operator.  With the term $H_{g} = 0$, this 
model is equivalent to a single particle version of our earlier many 
particle description, Eq.  (\ref{Eq.Heff}), and produces a time 
dependance for the probability of finding an atom in the cavity which 
is identical to $\langle a^{\dag}(t) a(t) \rangle$, found previously.  
These show a long time steady state in the number of atoms in the 
cavity which does not tend towards zero.  With the inclusion of 
gravity ($H_{g} \neq 0$), however, the number of atoms decays to zero 
in a non-exponential manner.  This behaviour is shown in Figure 
\ref{Fig.2}.

In figure \ref{Fig.2} we can see collapses and revivals in the number 
of atoms in the cavity.  This is due to the evolving phase 
relationship between the atoms in the cavity and the atoms which have 
been coupled into the output field modes.  There is no version of the 
Born-Markov approximation that can describe behaviour such as this, as 
such an approximation requires that there be no entanglement between 
the lasing mode and the output modes.
 
The presence of gravity changes the long time behaviour of our model 
so that the number of atoms asymptotes to zero, while the short time 
behaviour remains the same.  Other effects may also lead to the long 
time decay of atom number.  Repulsive interactions, for instance, 
would be expected to destroy the thin stable structure in position 
space which leads to the long time non-zero population of the cavity 
mode.  The effect of repulsive interactions can be included in this 
model by including in Eq.  (\ref{Eq.H2}) a nonlinear Hamiltonian term 
given by
\begin{eqnarray}
H_{i} &=& N U_{0} |\psi(x)|^{2}. \label{Eq.Hrep}
\end{eqnarray}
Including such interactions into our model produces a Gross-Pitaevskii 
type equation \cite{GPEqn}, where $N$ is the number of atoms in the 
system and $U_{0}$ an interaction strength.  We find that introducing 
such an interaction term does cause the atom number to go below the 
nonzero steady state predicted in the interaction free model.

\section{A non Markovian master equation} \label{sec:nonmarkov}

In the previous two sections we have shown that the standard master 
equation does not correctly describe the dynamics of our model for 
particular parameters.  However a master equation, in terms of only 
the system variables, would be an important tool for describing an 
atom laser.  We now consider whether a non-Markovian master equation 
can give a correct description of the atom laser.  To do this, we 
continue to make the Born approximation, but do not make a Markov 
approximation.

  The master equation with the Born approximation only is given in Eq.  
  (\ref{Eq.Born}).  Again, we check the validity of the Born 
  approximation by comparing the results obtained from this Born 
  master equation with the exact solutions.  We begin by considering 
  the resulting equation for the expectation value $\langle a^{\dag} 
  \rangle$, %
\begin{eqnarray}
\frac{d \langle a^{\dag}(t) \rangle}{dt} &=& i \omega_{0} \langle a^{\dag} 
\rangle (t) - \int_{0}^{t} dt' \langle a^{\dag}(t') \rangle \times 
\nonumber\\
&& \int |\kappa(k)|^{2} e^{i \omega_{k}(t-t')} 
dk,\label{Eq.dadagdtnm}
\end{eqnarray}

Eq.  (\ref{Eq.dadagdtnm}) is the same as that obtained through the 
full system plus reservoir equations given in Eq.  (\ref{Eq.adag}).  
This can be seen by making the transformation $\tau = t-t'$ in Eq.  
(\ref{Eq.dadagdtnm}) to obtain the alternative form, Eq.  
(\ref{Eq.adag}).  Despite this success, the density operator equation 
with the Born approximation, Eq.  (\ref{Eq.Born}) is not correct.  In 
particular, the Born approximation does not give correct values for 
higher order expectation values, such as $\langle a^{\dag} a \rangle$.

The equation derived from the non-Markovian master equation, Eq.  
(\ref{Eq.Born}) for the number operator expectation value is
\begin{eqnarray}
\frac{d \langle a^{\dag}(t) a(t) \rangle}{dt} &=& - \int_{0}^{t} 
d\tau \langle a^{\dag} a \rangle (t-\tau) \times \nonumber\\
&& \int |\kappa (k)|^{2} 
e^{i (\omega_{0} - \omega_{k}) \tau} dk + \mbox{h.c}, 
\label{Eq.dadagadtnm}
\end{eqnarray}
with solution
\begin{eqnarray}
\lefteqn{\langle a^{\dag}(t) a(t)\rangle = \langle a^{\dag}(0) a(0) 
\rangle  
\times } \hspace{1cm} \nonumber\\
& &  {\cal 
L}^{-1} \left\{ \frac{1}{s + {\cal L} \left\{f'(t) \right\}(s) + {\cal 
L} \left\{f'(t)^{*} \right\}(s)} \right\}(t).  \label{Eq.adagasolnnm}
\end{eqnarray}

\begin{figure}
\begin{center}
\epsfxsize=\columnwidth
\epsfbox{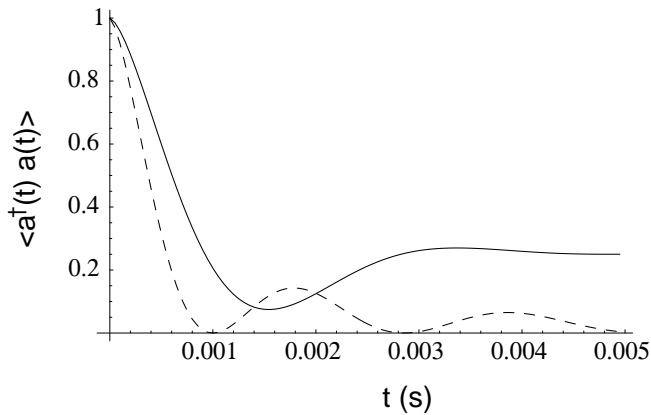}
\end{center}
\caption{A comparison between $\langle a^{\dag}(t) a(t) \rangle $ 
found using the Born (non-Markovian) master equation (dashed line) and 
the exact solution (solid line) respectively in the broadband regime.  
Atom numbers are normalized so that $\langle a^{\dag}(0) a(0) \rangle 
= 1$.  Parameters are the same as in Fig.  \ref{Fig.1}b.}
\label{Fig.3}
\end{figure}

These do not agree with the corresponding exact equations given in Eq.  
(\ref{Eq.adaga}) and Eq.  (\ref{Eq.adagasoln}), as is shown in Fig.  
\ref{Fig.3}.  
This figure compares the exact results for $\langle 
a^{\dag}(t) a(t) \rangle$ with the solution in the Born approximation 
only, Eq.  (\ref{Eq.adagasolnnm}).  The parameter values chosen are 
the same as those discussed in section \ref{sec:equations} and the 
comparison is made for broadband coupling.  The reason for the failure 
of the non-Markovian master equation is that where the correct 
equations lead to the term $\langle a^{\dag}(t) e^{-i \omega_{0} \tau} 
a(t-\tau) \rangle$, the non-Markovian master equation with the Born 
approximation leads to $\langle a^{\dag}(t-\tau) a(t-\tau) \rangle$.  
These two terms are equal if we consider only the free evolution of 
the system, and ignore the interaction term.  That is, to zeroth order 
in $|\kappa(k)|^{2}$, $\langle a^{\dag} (t-\tau) a(t-\tau) \rangle 
\approx \langle a^{\dag}(t) a(t-\tau) \rangle (t) e^{-i \omega_{0}t}$.  
However, there are interaction terms between the system and reservoir.  
Due to the interaction, terms obtained using the Born approximation 
are not equal to the exact results.  To reproduce equations for the 
correct system dynamics we cannot ignore the correlations which arise 
between the system and reservoir. 

In optics the Born-Markov approximation is a useful approximation in 
almost all practical parameter regimes.  In fact, in optics the value 
of $\omega_{0}$ is sufficiently large that, if one assumes a linear 
dispersion relation the reservoir correlation can be well 
approximatied by a Dirac delta function.  This is not the case for 
atoms.

The non-Markovian master equation, Eq.  (\ref{Eq.Born}), gives correct 
results for single operator expectation values.  Nevertheless, the 
Born approximation neglects higher orders in the interaction term, 
$|\kappa(k)|^{2}$, thus ignoring reservoir evolution.  This leads to 
incorrect results for higher order expectation values, such as 
$\langle a^{\dag}(t) a(t) \rangle$.

\section{Conclusion} \label{sec:Conclusion}

We have discussed the use of master equations and other density 
operator equations for atom lasers.  We find that the Born-Markov 
master equation, as is commonly used in optics, is not valid in 
parameter regimes in which atom lasers may work.  The output modes 
correlation function for photons is well approximated by a delta 
function in the broadband limit, and for typical optical parameters.  
In comparison, this function falls off as $1/\sqrt{\tau}$ for atoms 
because of the atom dispersion relations.

The Born-Markov approximations must be made self consistently.  In 
regimes where the Born-Markov approximations fail, the system can be 
solved in the Heisenberg picture, treating the output modes fully 
\cite{Hope97,Moy97b}.  For broadband coupling, the parameter regimes 
in which the Born-Markov approximation is valid is given in Eq.  
(\ref{Eq.inequality}).  Another possibility for describing such 
systems involves the use of quantum trajectories.  These have been 
found to be useful for other systems in which no Born-Markov master 
equation exists\cite{Bay97}.

The breakdown of the Born and Markov approximations means that new 
theoretical methods will be required to understand the atom laser in 
some regimes.  However, it also opens up the possibility of finding 
new properties of atom lasers significantly different from those found 
in the optical laser.

 %
 %

\end{document}